\documentstyle[aps,multicol,fleqn,epsfig,psfig]{revtex}

\begin{document}

\def\ra{\rangle}
\def\la{\langle}
\def\bege{\begin{equation}}
\def\ende{\end{equation}}
\def\begarr{\begin{eqnarray}}
\def\endarr{\end{eqnarray}}
\def\no{\noindent}
\def\non{\nonumber}
\def\hi{\hangindent=45pt}
\def\v{\vskip 12pt}
\def\Ap{{A^\prime}}
\def\Bp{{B^\prime}}
\def\ad{{{\hat a}^\dagger}}
\def\bd{{{\hat b}^\dagger}}
\def\cd{{{\hat c}^\dagger}}
\def\dd{{{\hat d}^\dagger}}

\newcommand{\ket} [1] {\vert #1 \rangle}
\newcommand{\bra} [1] {\langle #1 \vert}
\newcommand{\braket}[2]{\langle #1 | #2 \rangle}
\newcommand{\proj}[1]{\ket{#1}\!\!\bra{#1}}
\newcommand{\mean}[1]{\langle #1 \rangle}

\draft

\title{Linear optics and projective measurements alone suffice\\
to create
  large-photon-number path entanglement}

\author{Hwang Lee,$^1$ Pieter Kok,$^1$ Nicolas J.\ Cerf,$^{1,2}$ 
and Jonathan P.\ Dowling$^1$}

\address{$^1$
Quantum Computing Technologies Group, 
Exploration Systems Autonomy, Section 367 \\
 Jet Propulsion Laboratory,
 California Institute of Technology \\
 MS 126-347,
 4800 Oak Grove Drive, Pasadena, CA~~91109-8099\\
 $^2$ Ecole Polytechnique, CP 165, Universit\'e Libre de Bruxelles,
 1050 Brussels, Belgium
}

%\date{September 2001}
\date{\today}

\maketitle

\begin{abstract}
We propose a method for preparing maximal path entanglement with a
definite photon number $N$, larger than two, using projective
measurements. In contrast with the previously known schemes, our
method uses only linear optics. Specifically, we exhibit a way of 
generating four-photon, path-entangled states of the form 
$\ket{4,0}+\ket{0,4}$, using only four beam splitters and two
detectors. These states are of major interest as a resource for
quantum interferometric sensors as well as for optical quantum lithography
and quantum holography.
\end{abstract}


\begin{multicols}{2}

Quantum entanglement plays a central role in quantum communication 
and computation. It also provides a significant improvement
in frequency standards as well as in the performance 
of interferometric sensors\cite{wineland96,dowling98}. 
In this context, it has been shown that the Heisenberg limit
for phase sensitivity of a Mach-Zehnder interferometer can be reached by using 
maximally entangled states with a definite number of photons $N$, that is, 
%\bege  \label{N0+0N}
$
\ket{N,0}_{A,B}+\ket{0,N}_{A,B}.
$
%\ende
Here, $A$ and $B$ denote the two arms of the interferometer. These states, 
also called path-entangled photon number states, allow a phase
sensitivity of order $1/N$, whereas coherent light yields 
the shot-noise limit of $1/\sqrt{\bar{n}}$, with mean
photon number $\bar{n}$ \cite{scully97}.
The use of quantum entanglement can also be applied to optical
lithography. It has been shown recently that the Rayleigh diffraction
limit in optical lithography can be beaten by the use of
path-entangled photon number states\cite{boto00}.
In order to obtain an $N$-fold resolution enhancement, with quantum 
interferometric optical lithography, one again needs to create the $N$-photon 
path-entangled state given 
above.
%by Eq.\ (\ref{N0+0N}).
Due to interference
of the two paths, one obtains an intensity pattern at the lithographic 
surface which is proportional to $1+\cos N\varphi$, where $\varphi$ 
parametrizes the position on the surface. 
A 
%(incoherent)
superposition of these states with varying $N$ and suitable phase
shifts then yields a Fourier series of the desired pattern,
up to a constant\cite{kok01}. 
%% `pattern' instead of `image' because latter suggests 2D.

In view of these potential applications, finding methods 
for generating path-entangled states has been a longstanding endeavor
in quantum optics. 
Unfortunately, with the notable exception of $N=2$, 
the optical generation of these states seemed to require 
single-photon quantum logic gates that involve a large nonlinear 
interaction, 
namely, a Kerr element with $\chi^{(3)}$ on the order
of unity. Typically, $\chi^{(3)}$ is of the order $10^{-16}$ cm$^2$
s$^{-1}$ V$^{-2}$ \cite{boyd99}. 
This makes a physical implementation
with previously known techniques very 
difficult \cite{milburn89,franson99,gerry01}.
Recently, however, several methods for the realization
of probabilistic single-photon quantum logic gates have been proposed,
which make use solely of linear optics and projective measurements
(PMs) \cite{milburn01,imoto01,franson01}.
PMs are performed by 
measuring some part of the system while the rest of it is projected 
onto a desired state (state reduction). 
Since the state obtained is conditioned on a measurement outcome, 
this method only works probabilistically.
Such a protocol has been employed experimentally, by the
group of Zeilinger, to generate four-photon
{\it polarization} entanglement \cite{pan01}.

In this letter, we devise a technique for generating maximally
path-entangled photon number states based on this paradigm.
In particular, our method circumvents the use of $\chi^{(3)}$
nonlinearities in a Fredkin gate approach, for example \cite{gerry01}.  
We suggest several linear optical schemes,
based on projective measurements, 
for the preparation
of a four-photon, path-entangled state.
We also discuss the feasibility of these schemes, 
by investigating the consequence of inefficient detectors
on the state preparation process.

\begin{figure}[ht]
\epsfysize=2.0cm
\centerline{\epsfbox{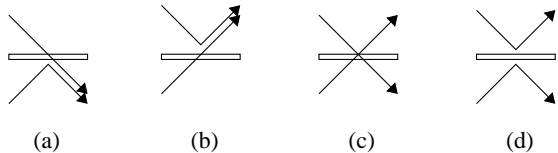}}
\v
\caption{
\label{fig-1}
Four possibilities when sending a $|1,1\ra$ state 
through a beam splitter. The diagrams (c) and (d) lead to the same final
state, but interfere destructively: (c) transmission-transmission $(i)(i)=-1$;
(d) refection-reflection $(-1)(-1)=1$.
}
\end{figure}

It is well known that two-photon,
path-entangled states can be created
using a Hong-Ou-Mandel (HOM) interferometer, 
where a photon pair from a parametric down-converter 
impinges onto a 50:50 beam splitter\cite{mandel87}. 
The beam splitter yields the path-entangled state 
$\ket{2,0}_{A'B'}+\ket{0,2}_{A'B'}$ from the product state 
$\ket{1,1}_{AB}$. In other words, the probability amplitude for
having $\ket{1,1}_{A'B'}$ at the output of the beam splitter vanishes. 
This can be understood by a simple diagrammatic analysis 
(see Fig.~\ref{fig-1}).

In our convention, the reflected mode acquires a phase $\pi$
while the transmitted mode acquires a phase of $\pi/2$,
consistent with the reciprocity requirement, 
so that the two possible ways of producing a state $\ket{1,1}$
interfere destructively \cite{dowling98b}. 
However, a beam splitter is not 
sufficient any more if the goal is to produce path-entangled states
with a photon number larger than two\cite{campos89}. 
Consequently, it is commonly assumed that $\chi^{(3)}$ 
nonlinear optical components are needed for $N >2$. 
By contrast, we show here that the recourse to such
nonlinearity can be avoided if single-photon detectors
are added to the scheme. 
The desired path-entangled states are then obtained, 
conditioned on the measurement outcome.

\narrowtext

\begin{figure}[ht]
\epsfysize=3.0cm
\centerline{\epsfbox{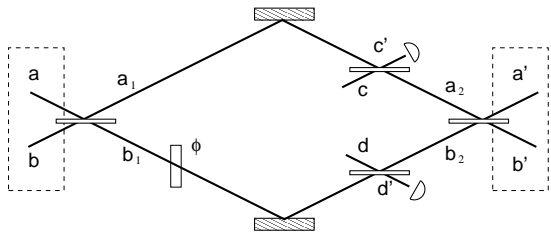}}
\v
\caption{ \label{fig-2} Mach-Zehnder interferometer with two
  additional beam splitters in the lower and upper arms, which direct
  the reflected beams to photodetectors. A one-photon count at both
  detectors allows the projective generation of the states
  $\ket{2,0}+\ket{0,2}$ or $\ket{4,0}+\ket{0,4}$, depending on the
  input state.}  
\end{figure}

Before considering the interesting case of $N=4$, 
it is instructive to first exhibit the generation of the state
$\ket{2,0}_{A'B'}+\ket{0,2}_{A'B'}$ using projective measurements,
instead of a simple beam splitter.
Let us consider a Mach-Zehnder interferometer with two
additional beam splitters, each of them being followed 
by a detector (see Fig.~\ref{fig-2}). 
In such a configuration, with all paths balanced, 
one can select the desired state via state reduction,
conditionally on both detectors clicking.
Formally, we are dealing with a four-port optical device, 
which may be characterized by expressing the output bosonic
mode operators
${\hat a}'$, ${\hat b}'$, ${\hat c}'$, and ${\hat d}'$
as a function of the input mode operators
${\hat a}$, ${\hat b}$, ${\hat c}$, and ${\hat d}$
\cite{dowling98}. 
For the transformation
effected by a single beam splitter 
(say, the first one in Fig.~\ref{fig-2}), 
we use the convention
%\begarr
$
{\hat a_1}=(-{\hat a}+ i {\hat b})/\sqrt{2}, 
~{\hat b_1}=(i{\hat a} - {\hat b})/\sqrt{2}
$.
%\endarr

Combining the transformations of the first, the last, and the two
intermediate beam splitters in the lower and upper arms,
we get the overall transformation
\begarr  
{\hat a'}&=&{\hat b}/\sqrt{2}+({\hat c}-i{\hat d})/2, \non\\
{\hat b'}&=&{\hat a}/\sqrt{2}+({\hat d}-i{\hat c})/2, \non\\
{\hat c'}&=&({\hat a}-i{\hat b})/2 + i {\hat c}/\sqrt{2}, 
\label{canonical} \\
{\hat d'}&=&({\hat b}-i{\hat a})/2 + i {\hat d}/\sqrt{2}. \non
\endarr
Note that we neglect here the phase induced by the mirrors and
that accumulated along the optical path, since they cancel for a
suitably balanced interferometer.
For a given input state, one obtains the output state
simply by expressing the input modes 
in terms of the output modes, 
that is, by inverting Eqs.~(\ref{canonical}).
Suppose the input state is 
$\ket{2,2}_{AB}={1\over 2} ({\hat a}^\dagger)^2 ({\hat b}^\dagger)^2 \ket{0}$.
Then, the term of order ${\hat c}^{\prime \dagger}{\hat d}^{\prime\dagger}$ 
in the expansion of 
$({\hat a}^\dagger)^2 ({\hat b}^\dagger)^2$ can be shown to be
$-{i \over 4}[ ({\hat a}^{\prime\dagger})^2+({\hat
  b}^{\prime\dagger})^2 ]$.
If we call $|\psi_{\rm am}\rangle$ the state before the
projective measurement ({\em ante} measurement), 
we have the output state ({\em post} measurement)
%\bege\label{amtopm2}
$
 |\psi_{\rm pm}\rangle = \langle 1,1|\psi_{\rm am}\rangle \propto 
 |2,0\rangle + |0,2\rangle\; .
$
%\ende
Thus, if one and only one photon is measured at each detector, 
one obtains the envisioned two-photon path-entangled output state.
The probability of this event is $1/16$.% $\approx 0.06$.

The way this projective method works can be understood very simply.
After passing through the first beam splitter, the product state
$|2,2\ra$ becomes a linear superposition of
$|4,0\ra$, $|2,2\ra$, and $|0,4\ra$.
Again, the states $|3,1\ra$ and $|1,3\ra$
do not appear for the same reason as the vanishing of
the HOM output state $|1,1\ra$, when the input is $|1,1\ra$ (see Fig~3).
Since the detection of one photon at each detector
requires at least one photon in both the upper
and lower arms of the interferometer,
the $|4,0\ra$ and $|0,4\ra$ states cannot contribute
to such events.  Consequently, only the $|2,2\ra$ term is left,
which then becomes $|1,1\ra$ if one photon is detected in each arm.
This $|1,1\ra$ state is thus found at the input of the last beam splitter, 
which results in the expected state $|2,0\ra+|0,2\ra$.

\begin{figure}[ht]
\epsfysize=4cm
\centerline{\epsfbox{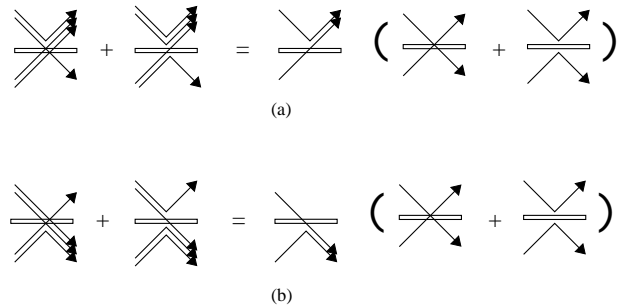}}
\v
\caption{
\label{fig-3}
Two possible ways of making (a) $|3,1\ra$ or (b) $|1,3\ra$ 
from an input state $|2,2\ra$ passing through a beam splitter. 
The two diagrams interfere destructively just as in Fig.~\ref{fig-1}.
}
\end{figure}

We can now use this approach to proceed to generate
the $|4,0\ra + |0,4\ra$ state. The key reason why projective measurement
is useful in the above scheme is that it enables us to conditionally suppress
the extreme components $\ket{4,0}$ and $\ket{0,4}$,
while leaving the middle component $\ket{2,2}$ unchanged. More generally,
the generation of path-entangled states with $N>2$ requires
eliminating the extreme components
with respect to the middle terms.
Suppose we want to produce the state $|4,0\ra + |0,4\ra$. 
Then, a simple matrix inversion shows that the state we need at the input
of the last beam splitter is generated an operator
of the form
$(\ad)^4 -6 (\ad)^2 (\bd)^2 + (\bd)^4$.
Similarly, to produce the output state $|4,0\ra - |0,4\ra$,
the required input operator is of the form
$(\ad)^3 (\bd)-(\ad) (\bd)^3$.
Since the latter operator has fewer terms, we will 
focus for the moment on producing $|4,0\ra - |0,4\ra$.

Let us show how to produce this state taking $|3,3\ra$ as the input state
and using the same interferometric setup as in Fig.~\ref{fig-2}.
The first beam splitter transforms 
$|3,3\ra ={1\over 6} (\ad)^3 (\bd)^3 \ket{0}$
into a linear superposition of
$|6,0\ra$, $|4,2\ra$, $|2,4\ra$, and $|0,6\ra$ generated by
\bege \label{bs33}
(\ad)^6 + 3(\ad)^4 (\bd)^2
+ 3 (\ad)^2 (\bd)^4 + (\bd)^6 
\ende
After passing through the two intermediate beam splitters,
and if one and only one photon is counted at each detector,
the state is then projected onto an equal superposition of $|3,1\ra$ and
$|1,3\ra$. 
Indeed, the states $|6,0\ra$ or $|0,6\ra$ are again eliminated 
by this projective measurement, 
since they cannot yield a click at {\em both} detectors.
The $|4,2\ra$ and $|2,4\ra$ states, on the other hand,
lose one photon in each arm of the interferometer and are therefore
reduced to $|3,1\ra$ and $|1,3\ra$, respectively.
Thus, just before the last beam splitter, we have
$|3,1\ra + |1,3\ra$. Finally, we need to add
a $\pi/2$-phase shifter in the lower arm of the
interferometer (see Fig.~\ref{fig-2}) 
in order to get the relative phase $\pi$ that is needed between
the two terms. This transforms Eq.~(\ref{bs33}) into
\bege 
(\ad)^6 - 3 (\ad)^4 (\bd)^2
+ 3 (\ad)^2 (\bd)^4 -  (\bd)^6 
\ende
so that the state after the projective measurement
is reduced to $|3,1\ra - |1,3\ra$. 
Consequently after the last beam splitter, we get the desired state
$|4,0\ra - |0,4\ra$. Of course, the state $|4,0\ra + |0,4\ra$
can simply be obtained by putting an extra
$\pi/4$-phase shifter at the end of one path.
A straightforward calculation shows that if the input state is
${1\over 6}(\ad)^3 (\bd)^3 |0\ra$. Then, as before,
the output state reads
%\bege\label{outstate33}
$ |\psi_{\rm pm}\rangle = \langle 1,1|\psi_{\rm am}\rangle \propto 
 |4,0\rangle + |0,4\rangle\; .
$
%\ende
A proper normalization shows that the probability to yield the desired 
state $|4,0\ra + |0,4\ra$ is $3/64$. % $ \approx 0.05$.
Note that any $|2N+1,2N+1\ra$ input state may be used 
in this configuration to yield $|4,0\ra + |0,4\ra$  
by detecting  $2N-1$ photons at each detector,
but with a smaller yield as $N$ increases.

An alternative way of producing $|4,0\ra + |0,4\ra$
was found that requires the ability of preparing the input states 
$\ket{2,2}$ and $\ket{1,1}$, instead of $\ket{3,3}$. 
The idea is to feed the previously unused input ports of 
the two intermediate beam splitters (modes $c$ and $d$ in Fig.~\ref{fig-2})
with the state $|2,0\ra + |0,2\ra$.
This state is obtained
by sending $\ket{1,1}$ through a HOM beam splitter.
Suppose we have an input state $|2,2\ra$, which
after the first beam splitter gives a superposition of
$|4,0\ra$, $|2,2\ra$, and $|0,4\ra$, as explained above.
Consider, first, the middle term $|2,2\ra_{A_1B_1}$, which gives
\begarr
\lefteqn{ \ket{2,2}_{A_1 B_1} (\ket{2,0}_{CD} + \ket{0,2}_{CD})  } \non\\
&=&\ket{2,2}_{A_1 C} \ket{2,0}_{B_1 D} + \ket{2,0}_{A_1 C} \ket{2,2}_{B_1 D} ,
\endarr
so that either the beam splitter in the upper arm or that in the lower
arm is fed again with $\ket{2,2}$. As shown in Fig.~\ref{fig-3}, 
this leads to the measurement of zero or two photons at the
corresponding detector, but cannot give one count. 
Consequently, 
the middle term cannot contribute to $|1,1\ra_{C'D'}$. 
Take now the first term $|4,0\ra_{A_1 B_1}$, which gives
\begarr
\lefteqn{ |4,0\ra_{A_1 B_1} (|2,0\ra_{CD} + |0,2\ra_{CD}) }
\non \\
&=& |4,2\ra_{A_1 C} |0,0\ra_{B_1 D}+ |4,0\ra_{A_1 C} |0,2\ra_{B_1 D}.
\endarr
Clearly, the first term in the latter expression
cannot give a click at the lower detector. 
In contrast, the second term can give a click at both detectors, 
which results in
the state $|3,1\ra_{A_2 C'} |1,1\ra_{B_2 D'}$, 
after the intermediate beam splitters. 
Thus, postselecting on one count
at each detector yields $|3,1\ra_{A_2 B_2}$.
Similarly, for the third term $|0,4\ra_{A_1 B_1}$, we get
the state $|1,3\ra_{A_2 B_2}$ after postselection.
Consequently, we only now need to adjust the relative
phase between the $|4,0\ra_{A_1B_1}$ and $|0,4\ra_{A_1B_1}$ states
in order to get $\ket{3,1}-\ket{1,3}$ before the last beam splitter.
This can be done by inserting a $\pi/4$-phase shifter
in the lower arm of the interferometer. 
Then the desired state $|4,0\ra - |0,4\ra$ is produced after 
the last beam splitter.
A simple calculation shows that an input state
${1\over 4}(\ad)^2 (\bd)^2 [(\cd)^2 + (\dd)^2] |0\ra$
yields the same output as before
%in Eq.~(\ref{outstate33})
up to an irrelevant global phase,
%\begarr
%\cdots + {i \over 32} \big[ (\ad)^4 - (\bd)^4 \big] |0,0\ra_{A,B}
%|1,1\ra_{C,D} +\cdots
%\endarr
so that one-photon detection at each detector projects the output state onto 
$|4,0\ra -|0,4\ra$ with probability $3/64$. 
The yield is thus equal to
that of the previous scheme.
Note again, that with this configuration,
any $|2N,2N\ra(|2,0\ra-|0,2\ra)$ input state yields,
conditionally on the detection of $2N-1$ photons at each detector,
the same output state $|4,0\ra - |0,4\ra$.
However, the probabilities decrease as $N$ increases.

The schemes we have shown so far
relied on symmetric product states $|N,N\ra$ as inputs. States of this
form are typically produced in optical parametric oscillators 
and down-converters \cite{kim98,kok01b}.
We have also devised schemes, which 
start from the state $|5,0\ra$ instead,
and from which we generate states of the form 
$|N,0\ra + |0,N\ra$, for $N \in \{2,3,4\}$ (see Fig.~4).
Such input states as $|N,0\ra$ can be produced by
manipulating states of the form $|N,N\ra$,
or from $N$-photon sources,
now under development \cite{kok01b,kim}.

\begin{figure}[ht]
\epsfysize=3.0cm
\centerline{\epsfbox{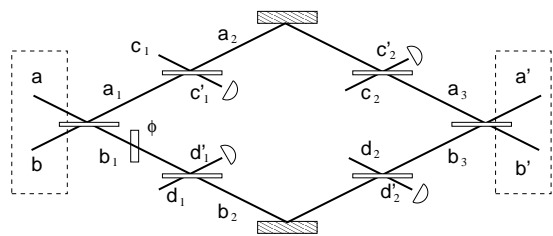}}
\v
\caption{
\label{fig-4}
Four-detector scheme with a
Mach-Zehnder interferometer.
When we feed the $c_1, d_1$ modes
with $|N-1,0\ra_{C_1D_1}+|0,N-1\ra_{C_1D_1}$,
we have the output state
$|N,0\ra_{A'B'}-|0,N\ra_{A'B'}$
($N\in\{2,3,4\}$) conditioned upon
one photon detection at each detector.
Here the input state is
$|5,0\ra_{AB}$ and $\phi=\pi/2$.
}
\end{figure}

%
% Inefficient detection
%

Finally, let us discuss the consequence of using realistic detectors
in our schemes.  We can model the detector efficiency $\eta^2$ 
with an ideal detector preceded by a beam splitter 
with transmissivity $\eta$. The photons deflected from the detector 
represent the loss. When two photons
enter the inefficient detector, one of them might be lost, thus
yielding an incorrect detector outcome. This is particularly important
here, since we condition the outgoing state on single-photon
detection events. The projective measurement associated with
a single-photon detection can be modeled by the projector
%\begin{eqnarray}
$  \sum_{n=1}^{\infty} n\, \eta^2(1-\eta^2)^{n-1} |n\rangle\langle n|\; .
$
%\end{eqnarray}
Applying this to the first proposed scheme for generating 
$|4,0\rangle_{A'B'} - |0,4\rangle_{A'B'}$  (see Fig.~\ref{fig-2}), 
we obtain a state $\rho_{A'B'} \propto
\sum_{n,m=1}^{6} nm\eta^4 (1-\eta^2)^{n+m-2} \rho^{(n,m)}_{A'B'}$,
where $n,m$ are the number of photons lost in modes $C'$ and $D'$, 
and $\rho^{(n,m)}_{A'B'} \propto 
~_{C'D'}\!\langle n,m|\rho_{A'B'C'D'}|n,m\rangle_{C'D'}$. 
These density matrices $\rho^{(n,m)}_{A'B'}$, which arise 
due to imperfect detections, also correspond to $N$-photon 
path-entangled states, but with $N<4$.
\begin{center}
\begin{tabular}{c|cc}
  $(n,m)$ & $\quad$ & $\rho^{(n,m)}$ \cr \hline 
  $(1,2)$\rule{0pt}{12pt} & & $|3,0\rangle + |0,3\rangle$ \cr
  $(2,2)$ & & $|2,0\rangle + |0,2\rangle$ \cr
  $(3,1)$ & & $|2,0\rangle + |0,2\rangle$ \cr
  $(3,2)$ & & $|1,0\rangle + |0,1\rangle$ \cr
  $(4,1)$ & & $|1,0\rangle + |0,1\rangle$ \cr
\end{tabular}
\end{center}
%\begin{center}
%  $(n,m)$ & $\quad \quad$ & $\rho^{(n,m)}$ \cr 
%
%------------------------------------------
%
%  ~~$(1,2)$\rule{0pt}{12pt} & & ~~$|3,0\rangle + |0,3\rangle$ \cr
%
%  ~~$(2,2)$ & & ~~$|2,0\rangle + |0,2\rangle$ \cr
%
%  ~~$(3,1)$ & & ~~$|2,0\rangle + |0,2\rangle$ \cr
%
%  ~~$(3,2)$ & & ~~$|1,0\rangle + |0,1\rangle$ \cr
%
%  ~~$(4,1)$ & & ~~$|1,0\rangle + |0,1\rangle$ \cr
%\end{center}
%~

\no
{\small Table. The outgoing states $\rho^{(n,m)}$ of
the interferometer of Fig.~2 (only the ket parts are given since these
states are pure). The left column lists the photon-number coincidence in
the two detectors, while the right column gives the corresponding
outgoing state. When the detector outcomes are interchanged, i.e.,
$(n,m)\rightarrow(m,n)$, the corresponding state picks up a relative
minus sign.}\bigskip

Thus, the output state is a mixture of path-entangled states with
different values of $N$. 
For a realistic, single-photon resolution, photo-detector with 
efficiency $\eta^2=0.88$ \cite{kim}, the fidelity of the outgoing state
with respect to the envisioned state 
$|\Psi\rangle = |4,0\rangle +  |0,4\rangle$ is
%\begin{equation}
$
  F = \langle\Psi|\rho|\Psi\rangle = 0.64,
$
conditioned on a single-photon detector coincidence.
%\end{equation}
Even though these imperfect detections lead to a degraded fidelity,
this might be exploited in order to create incoherent superpositions of 
path-entangled states, which may be useful for the pseudo-Fourier 
method in quantum lithography\cite{kok01}.

%Recall that the state $|N,0\rangle +
%|0,N\rangle$ gives rise to a deposition rate $1 + \cos(N\varphi)$ on
%the substrate. Superposing these patterns with suitable intensities
%for different $N$ yields a Fourier series up to a constant. This is 
%useful for the pseudo-Fourier method in quantum lithography
%\cite{kok01}. Note that we do not need a coherent superposition of these 
%states since there is no interference anyway.

%Conclusion

In conclusion,
we have shown that conditioning the output of a linear optical setup
on single-photon detection events makes it possible to generate 
path-entangled photon number states with more than two photons. 
The price of eliminating nonlinear components
is the relatively low yield of the projective process, 
which is only about 5\% for the state $\ket{4,0}+\ket{0,4}$. 
Of course, the optical schemes we have found so far
are not necessarily the most efficient ones, so finding the
optimal protocols remains an interesting open problem. 
In particular, employing the teleportation ``fix'' used by Knill,
Laflamme and Milburn \cite{milburn01},
in future work we plan to
% it seems possible to 
devise schemes where the yield scales more
efficiently with $N$.

Another inherent difficulty is that our proposed schemes require
detectors that are able to resolve one or more photons. 
This problem may, however, not be critical 
in applications where incoherent superpositions of path-entangled 
photon number states are needed anyway, such as in quantum lithography. 
The projective generation method also requires the availability of
photon-number sources, which clearly is another challenge \cite{walther01}.
Finally, even though it is very likely that the technique presented here 
could be extended to generating path-entangled states 
with arbitrary $N$, we currently do not know of a generation scheme 
with $N>4$. Such schemes would probably require more complicated
interferometers with more detectors and tunable components.
This will be the subject of further investigation.

%This work was carried out by the Jet Propulsion Laboratory,
%California Institute of Technology, 
%under a contract with the National Aeronautics
%and Space Administration.
We wish to thank P.\ G.\ Kwiat, D.\ J.\ Wineland, Y.\ H.\ Shih,
J.\ D.\ Franson, C.\ Adami, G.\ M.\ Hockney, D.\ V. Strekalov,
C.\ P.\ Williams, and U.\ H.\ Yurtsever for useful
discussions. 
We would also like to acknowledge support from  NASA, ONR
and ARDA.
In addition, H.L.\ and P.K.\ would like to acknowledge
the National Research Council.
%Part of this work was performed while HL and PK
%held National Research Council Research Associateship
%Awards at the Jet Propulsion Laboratory.

\end{multicols}

\end{document}